\documentclass[aps,prb,twocolumn,groupedaddress,showpacs]{revtex4}

\usepackage{latexsym,amssymb,amsmath,amsfonts}
\usepackage{graphicx,color,pstricks}
\usepackage{epsfig,epsf,bm}

\definecolor{gray}{rgb}{0.7,0.7,0.7}

\begin{document}

\title{Electrical transport through a quantum dot side-coupled to a topological superconductor}

\author{Yu-Li Lee}
\email{yllee@cc.ncue.edu.tw} \affiliation{Department of Physics, National Changhua University of Education,
Changhua, Taiwan, Republic of China}

\date{\today}

\begin{abstract}
 We propose to measure the differential conductance $G$ as a function of the bias $V$ for a quantum dot
 side-coupled to a topological superconductor to detect the existence of the chiral Majorana edge states. It
 turns out that $G$ for the spinless dot is an oscillatory (but not periodic) function of $eV$ due to the
 coupling to the chiral Majorana edge states, where $-e$ is the charge carried by the electron. The behavior of
 $G$ versus $eV$ is distinguished from the one for a multi-level dot in three respects. First of all, due to the
 coupling to the topological superconductor, the value of $G$ will shift upon adding or removing a vortex in the
 topological superconductor. Next, for an off-resonance dot, the conductance peak in the present case takes a
 universal value $e^2/(2h)$ when the two leads are symmetrically coupled to the dot. Finally, for a symmetric
 setup and an on-resonance dot, the conductance peak will approach the same universal value $e^2/(2h)$ at large
 bias.
\end{abstract}

\pacs{
 73.63.-b 
 73.21.-b 
 74.45.+c 
}

\maketitle

\section{Introduction}

Recently, to search for the topological phases and to study their properties become crucial issues in condensed
matter physics. Among these topological matters, Majorana fermions, which were theoretically predicted to exist at
the edge or the core of a vortex in a $p_x+ip_y$-wave superconductor and superfluid,\cite{Read-Green,Gurarie,MStone}
attract a lot of attentions. This is largely triggered by the fact that Majorana fermions are stable against local
perturbations\cite{kitaev2} and obey the non-Abelian statistics,\cite{Ivanov} so that they have great potential in
the applications of fault-tolerant quantum computations.\cite{Kitave,Nayak}

Up to now, there are two main theoretical proposals to realize Majorana fermions. One way is to generate a
localized Majorana mode at the end of a spin-orbit coupled nanowire subjected to a magnetic field and proximate
to an $s$-wave superconductor,\cite{Lutchyn-Sau-DasSarma,Oreg-Refael-Open} which is motivated by a model proposed
by Kitaev.\cite{kitaev2} Experimental evidence for such a Majorana edge mode was obtained in indium antimonide
quantum wires.\cite{Mourik,nano} Majorana fermions can also be supported in the vortices of $s$-wave superconductors
deposited on the surface of a three-dimensional topological insulator.\cite{Fu-Kane} In particular, chiral Majorana
edge states can be created at the interface between a superconductor and the area gapped by ferromagnetic materials.\cite{Fu-Kane}

One of the challenges in the physics of Majorana fermions is to detect and to verify the existence of Majorana
fermions. Several methods have been proposed, including the noise measurement,\cite{BD,NAB} resonant Andreev
reflection,\cite{KTLaw} and the $4\pi$-periodic Josephson effect.\cite{Lutchyn-Sau-DasSarma,Oreg-Refael-Open,Fu-Kane2,Law,IF}
Recently, several groups proposed to measure the electric transport through a quantum dot (QD) coupled to the end
of a one-dimensional ($1$D) topological superconductor (TSC) to detect the existence of Majorana modes\cite{LF,LB,Vernek}
as well as their dynamics.\cite{Cao} Especially, in the case of a spinless QD side-coupled to the end of a TSC, the
zero-bias conductance at zero temperature takes the value $e^2/(2h)$,\cite{LB,Vernek} instead of $e^2/h$. This
fact has been identified as the evidence of the existence of Majorana end modes. In the present work, we propose
to employ the similar idea to detect the chiral Majorana edge states in the TSC.

\begin{figure}
\begin{center}
 \includegraphics[width=0.9\columnwidth]{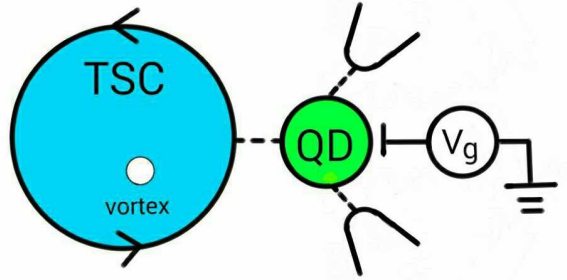}
 \caption{(Color online) A schematic setup of a QD side-coupled to a TSC and to two metallic leads. The dot level
 can be controlled by a capacitively coupled gate voltage $V_g$. The bias is applied between two leads. The TSC is
 formed by depositing a superconducting island on a three-dimensional topological insulator. The area outside the
 superconductor is gapped by ferromagnetic materials. At the interface between the superconductor and the
 ferromagnetic material, there is a branch of chiral Majorana fermions denoted by the arrow.}
 \label{set}
\end{center}
\end{figure}

A schematic setup is shown in Fig. \ref{set}. The QD, which can be formed by using either graphene\cite{grap} or 
the carbon nanotube,\cite{cnt} is side-coupled to a TSC and to two metallic leads. The dot level can be controlled 
by a capacitively coupled gate voltage $V_g$. The bias $V$ is applied between two leads. The TSC is formed by 
depositing a superconducting island on a three-dimensional ($3$D) topological insulator. (The possible candidates 
of $3$D topological insulators are Bi$_2$Se$_3$ or Bi$_2$Te$_3$.\cite{3DTI}) The area outside the superconductor is 
gapped by ferromagnetic materials. At the interface between the superconductor and the ferromagnetic material, there 
is a branch of chiral Majorana fermions. The presence of the ferromagnetic material removes spin degeneracy of the 
dot levels. Suppose that the Zeeman splitting is large enough. We may assume that the dot electrons are spinless.\cite{foot1} 
For such a case, we found that the differential conductance $G$ through the spinless dot is an oscillatory (but 
not periodic) function of $eV$ due to the coupling to the chiral Majorana edge states, where $-e$ is the charge 
carried by the electron. The behavior of $G$ versus $eV$ in the present case is distinguished from the one for a 
multi-level dot in three respects. First of all, the value of $G$ for the former will shift upon adding or removing 
a vortex in the TSC. Next, for an off-resonance dot, the conductance peak for the former takes a universal value 
$e^2/(2h)$ when the two leads are symmetrically coupled to the dot. Finally, for a symmetric setup and an 
on-resonance dot, the conductance peak in the present case will approach the same universal value $e^2/(2h)$ at 
large bias.

The rest of the paper is organized as follows. We first write down the Hamiltonian which models the setup in Fig.
\ref{set} and discuss the approximations we made in the calculations. Next, we present the relevant Green
functions to calculate the current through the dot. Then, we give the spectral function of the dot electrons, the
differential conductance, and the relevant stuffs. The final section is devoted to a summary of our results.

\section{The Model}

We consider a setup shown in Fig. \ref{set}. The TSC is realized by depositing a superconducting island on the
surface of a three-dimensional topological insulator. The region outside the superconductor is gapped by
ferromagnetic materials. This system can be modeled by the Hamiltonian: $H=H_C+H_D+H_T$, where
\begin{equation}
 H_C=\! \sum_{\alpha\in L,R} \! \sum_{\bm{k}}\epsilon_{\bm{k}\alpha}c^{\dagger}_{\bm{k}\alpha}
 c_{\bm{k}\alpha} \ , \label{mfqdh1}
\end{equation}
describes the leads, and
\begin{equation}
 H_T=\! \sum_{\alpha\in L,R} \! \sum_{\bm{k}}\frac{V_{\bm{k}\alpha}}{\sqrt{\Omega}}c^{\dagger}_{\bm{k}\alpha}d
 +\mathrm{H.c.} \ , \label{mfqdh11}
\end{equation}
describes the tunneling between the leads and the dot. The presence of the ferromagnetic material will split the
electronic levels of the dot with different spins. We shall focus on the case with large Zeeman splitting such
that within the energy scale in which we are interested the dot electrons can be regarded as spinless fermions.
We further assume that $eV$ and $k_BT$ are much smaller than the superconducting gap in the TSC and the average
level spacing in the dot, where $T$ is the temperature. Within these approximations, the Hamiltonian $H_D$ can be
written as
\begin{equation}
 H_D=\epsilon_dd^{\dagger}d+\frac{v_M}{2} \! \! \int^{L/2}_{-L/2} \! dx\eta(-i\partial_x\eta)-i\bar{t}\eta(0)
 (\xi d+\mathrm{H.c.}) \ , \label{dh1}
\end{equation}
where $\bar{t}>0$ denotes the tunneling amplitude between the dot and the Majorana edge states, $\xi$ is a
complex number with $|\xi|=1$, $|v_M|$ is the speed of Majorana fermions, and $L$ is the circumference of the
island. In Eq. (\ref{dh1}), we have taken $x=0$ as the contact point between the TSC and the dot. The dot level
$\epsilon_d$ can be adjusted by a capacitively coupled gate voltage $V_g$.

The real field $\eta(x)$, which describes the chiral Majorana edge states, obeys the anticommutation relation
\begin{eqnarray*}
 \{\eta(x),\eta(y)\}=\delta(x-y) \ .
\end{eqnarray*}
The Fourier decomposition of $\eta(x)$ is given by
\begin{eqnarray*}
 \eta(x)=\frac{1}{\sqrt{L}}\sum_k\psi_ke^{ikx} \ ,
\end{eqnarray*}
where $\psi_k$ and $\psi_k^{\dagger}$ obey the canonical anticommutation relations. Since $\eta$ is a real
field, we have $\psi_{-k}=\psi^{\dagger}_k$. The boundary condition of $\eta(x)$ depends on the number of vortices
$N_v$ in the TSC:
\begin{eqnarray*}
 \eta(L/2)=(-1)^{N_v+1}\eta(-L/2) \ ,
\end{eqnarray*}
which leads to $k=\frac{(2n+N_v+1)\pi}{L}$ with $n=0,\pm 1,\pm 2,\cdots$.

For simplicity, we assume that $V_{\bm{k}\alpha}$ is independent of $\bm{k}$, leading to the level-width functions
$\Gamma_{L(R)}(E)=2\pi |V_{L(R)}|^2N_{L(R)}(E)$, where $N_{L(R)}(E)$ is the density of states for electrons in the
left (right) lead. We further ignore the energy dependence of $\Gamma_{L(R)}(E)$. Within these approximations, the
current can be calculated by a Landauer-type formula:\cite{MW}
\begin{equation}
 I=-\frac{e\Gamma_L\Gamma_R}{2\pi\Gamma} \! \int^{+\infty}_{-\infty} \! dE[f_L(E)-f_R(E)]A(E) \ , \label{i1}
\end{equation}
where $\Gamma=\Gamma_L+\Gamma_R$ and $f_{L(R)}(E)=[e^{\beta (E-\mu_L)}+1]^{-1}$ is the distribution function of
electrons in the left (right) lead with $\mu_{L(R)}$ being the corresponding chemical potential and
$\beta =(k_BT)^{-1}$. We shall take $\mu_L-\mu_R=-eV$. Moreover, we set $\mu_L=0=\mu_R$ in equilibrium. In Eq.
(\ref{i1}), $A(E)=-2\mbox{Im}[D_r(E)]$ is the spectral function for dot electrons and $D_r(E)$ is the Fourier
transform of the retarded Green function for dot electrons:
\begin{eqnarray*}
 D_r(t_1,t_1)=-i\Theta(t_1-t_2)\langle\{d(t_1),d^{\dagger}(t_2)\}\rangle \ .
\end{eqnarray*}
By taking $\mu_L=-eV/2$ and $\mu_R=eV/2$, we find that
\begin{equation}
 G(V)=\frac{dI}{dV}=\frac{\Gamma_L\Gamma_R}{\Gamma} \! \left[\frac{A(eV/2)+A(-eV/2)}{2}\right] \! G_0 \ ,
 \label{i11}
\end{equation}
at $T=0$, where $G_0=e^2/(2\pi)$ is the conductance quantum for spinless electrons. Equation (\ref{i11}) indicates
that $G(V)$ measures the symmetric part of $A(E)$. Hence, the rest of the task is to calculate $A(E)$.

\section{Electrical transport through a spinless dot}
\subsection{The local density of states for dot electrons}

\begin{figure}
\begin{center}
 \includegraphics[width=0.9\columnwidth]{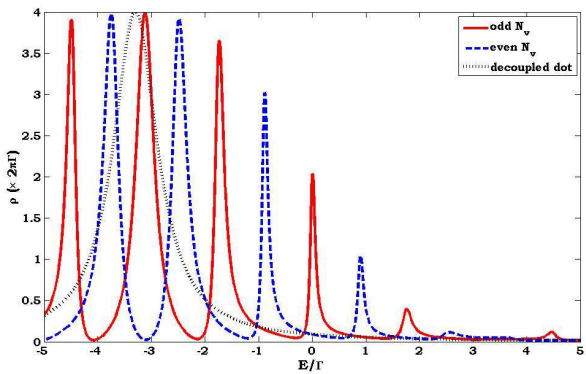}
 \caption{(Color online) LDOS's for the dot. Solid (dashed) line represents the case with an odd (even) number of
 vortices in the superconductor. We take $\Gamma_L=\Gamma_R=\Gamma/2=0.5\pi v_M/L$, $\epsilon_d=-3.3\Gamma$, and
 $\bar{t}^2/v_M=2\Gamma$. The location of the peak will shift when a vortex is added or removed. For comparison,
 the dotted line corresponds to the case with $\bar{t}=0$.}
 \label{dos1}
\end{center}
\end{figure}

\begin{figure}
\begin{center}
 \includegraphics[width=0.9\columnwidth]{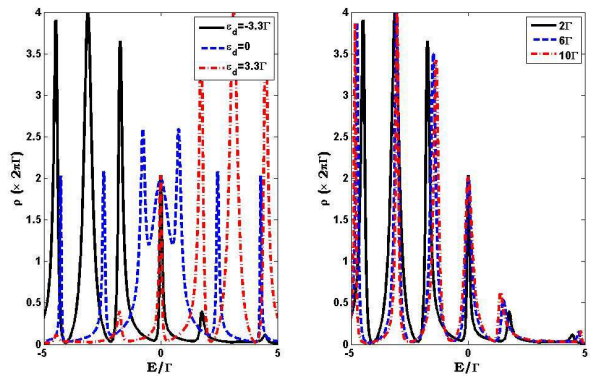}
 \caption{(Color online) LDOS's for the dot with odd $N_v$. We take $\Gamma_L=\Gamma_R=\Gamma/2=0.5\pi v_M/L$.
 {\bf Left}: $\rho(E)$ with $\bar{t}^2/v_M=2\Gamma$ for different values of $\epsilon_d$. The location of the
 resonance depends on the value of $\epsilon_d$ sensitively. {\bf Right}: $\rho(E)$ with $\epsilon_d=-3.3\Gamma$
 for different values of $\bar{t}^2/v_M$. The location of the resonance also slightly depends on the value of
 $\bar{t}^2/v_M$.}
 \label{dos2}
\end{center}
\end{figure}

One way to calculate $D_r(E)$ is to employ the method of equations of motion (EOM). Within the approximation we
have made, the set of EOM's for two-point Green functions is closed and one may get an exact form of $D_r(E)$:
\begin{equation}
 D_r(E)=\frac{E+\epsilon_d-M(E)+\frac{i}{2}\Gamma}
 {\left[E-M(E)+\frac{i}{2}\Gamma\right]^2 \! -\epsilon_d^2-[M(E)]^2} \ , \label{mdgf1}
\end{equation}
where
\begin{eqnarray*}
 M(E)=-\frac{\bar{t}^2}{2v_M}\tan{\! \left(\frac{L}{2v_M}E-\frac{N_v\pi}{2}\right)} ,
\end{eqnarray*}
arises from the propagator of Majorana fermions. In terms of Eq. (\ref{mdgf1}), the spectral function of the dot
electrons takes the form
\begin{equation}
 A(E)\! =\! \frac{\Gamma\{[E+\epsilon_d-M(E)]^2+[M(E)]^2+\frac{\Gamma^2}{4}\}}
 {\left[E^2\! -\! 2EM(E)\! -\epsilon_d^2\! +\! \frac{\Gamma^2}{4}\right]^2 \! \! +\! \Gamma^2 \! \left\{\epsilon_d^2\! + \! [M(E)]^2\right\}}
 . \label{mdgf11}
\end{equation}
The behavior of the local density of states (LDOS) of the dot, $\rho(E)=A(E)/(2\pi)$, is shown in Figs.
\ref{dos1} and \ref{dos2}.

A few remarks about $\rho(E)$ are in order. First of all, we notice that $\rho(-E,\epsilon_d)=\rho(E,-\epsilon_d)$.
Consequently, $\rho(-E)=\rho(E)$ when $\epsilon_d=0$, as shown in the left diagram in Fig. \ref{dos2}. Next,
$\rho(E)$ (or $A(E)$) is an oscillator function of $E$ on account of the coupling to the quantized energy levels of
the chiral Majorana edge states. The location of the peak indicates the occurrence of a resonance, which depends on
the number of vortices $N_v$ as well as the value of $\epsilon_d$. Variation of the value of $\bar{t}^2/v_M$ also
slightly shifts the position of the resonance, but does not change the global feature significantly.  For
$\epsilon_d<0$ ($\epsilon_d>0$), most resonances have energies $E<0$ ($E>0$). Especially, a zero-energy resonance
always exists when $N_v$ is odd, irrespective of the values of $\epsilon_d$ and $\bar{t}^2/v_M$. On the other hand,
for an off-resonance dot, there is at least a resonance lying between $\epsilon_d$ and $0$ when $N_v$ is even.

\subsection{The differential conductance}

Inserting Eq. (\ref{mdgf11}) into Eq. (\ref{i11}), we obtain the differential conductance at zero temperature. The
behavior of $G$ versus $eV$ is shown in Figs. \ref{cond1} -- {\ref{cond3}. We see that $G$ depends on $N_v$. When
$N_v$ is odd, a zero Majorana edge mode exists, which results in a zero-bias peak with the value
$G/G_0=2\Gamma_L\Gamma_R/\Gamma^2$, even for an off-resonance dot. As noticed in the previous work on the
topological superconducting nanowire,\cite{LB,Vernek} this is an indication of the existence of a zero-energy
Majorana edge mode. For even $N_v$ such that no zero-energy edge modes exist in the TSC, the zero-bias conductance
reaches the unitary value $G/G_0=4\Gamma_L\Gamma_R/\Gamma^2$ for an on-resonance dot. For an off-resonance dot, the
conductance exhibits an oscillatory behavior as varying $eV$, and reaches half of the value in the unitary limit at
the peak. Moreover, there is at least a conductance peak occurring at a value of $eV$ smaller than $2|\epsilon_d|$.
These provide evidences for the coupling to quantized energy levels of chiral Majorana edge states. The positions
of the conductance peaks shift by varying the gate voltage $V_g$, but are insensitive to the value of $\bar{t}^2/v_M$.
For an on-resonance dot, i.e. $\epsilon_d=0$, a conductance peak with the value larger than
$G/G_0=2\Gamma_L\Gamma_R/\Gamma^2$ occurs at a finite bias due to the enhanced side peaks in $\rho(E)$ (or $A(E)$)
with odd $N_v$, as shown in the left diagram in Fig. \ref{dos2}. To sum up, the whole behavior of $G$ with varying
$N_v$ and $eV$ as we have discussed is an indication of the existence of a chiral Majorana liquid at the edge of a
TSC.

One may wonder how to distinguish the behavior of $G$ versus $eV$ for a dot side-coupled to a TSC from that for a
multi-level dot. According to the above results, both are different in three respects. First of all, the
conductance for a dot side-coupled to a TSC will shift upon adding or removing a vortex in the TSC, while the one
for a dot decoupled to the TSC remains intact. Next, for an off-resonance dot, the value of the conductance peak
in the present case is only half of the one in the unitary limit, whereas for a dot decoupled to the TSC, it will
reach a non-universal value depending on the energy levels in the dot as long as $eV/2$ matches one level in the
dot. Finally, for an on-resonance dot, the conductance peak in the present case will approach a universal value
$G/G_0=2\Gamma_L\Gamma_R/\Gamma^2$ at large bias. In general, there is no such a behavior for a dot decoupled to
the TSC.

\begin{figure}
\begin{center}
 \includegraphics[width=0.9\columnwidth]{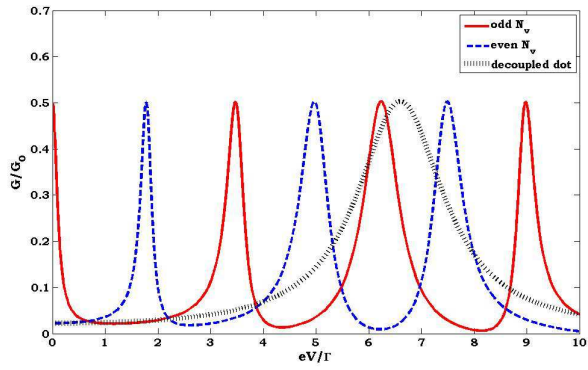}
 \caption{(Color online) The conductance (in units of $G_0$) versus $eV$ (in units of $\Gamma$). Solid (dashed) line
 represents the case with an odd (even) number of vortices in the superconductor. We take
 $\Gamma_L=\Gamma_R=\Gamma/2=0.5\pi v_M/L$, $\epsilon_d=-3.3\Gamma$, and $\bar{t}^2/v_M=2\Gamma$. For comparison,
 the dotted line corresponds to the case with $\bar{t}=0$.}
 \label{cond1}
\end{center}
\end{figure}

\begin{figure}
\begin{center}
 \includegraphics[width=0.9\columnwidth]{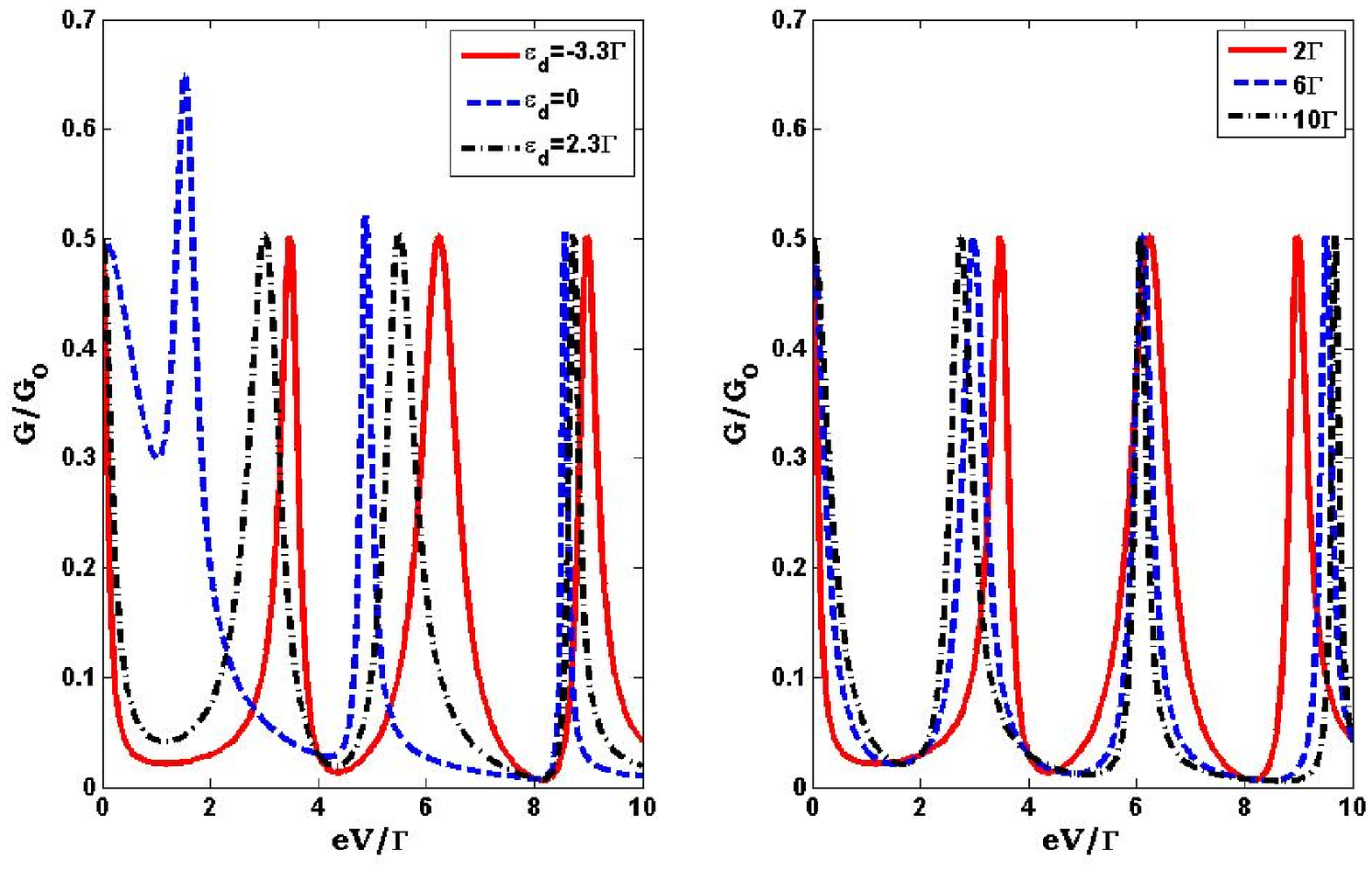}
 \caption{(Color online) The conductance (in units of $G_0$) versus $eV$ (in units of $\Gamma$) with odd $N_v$.
 We take $\Gamma_L=\Gamma_R=\Gamma/2=0.5\pi v_M/L$. {\bf Left}: $G$ as a function of $eV$ with $\bar{t}^2/v_M=2\Gamma$
 for different values of $\epsilon_d$. {\bf Right}: $G$ as a function of $eV$ with $\epsilon_d=-3.3\Gamma$ for
 different values of $\bar{t}^2/v_M$.}
 \label{cond2}
\end{center}
\end{figure}

\begin{figure}
\begin{center}
 \includegraphics[width=0.9\columnwidth]{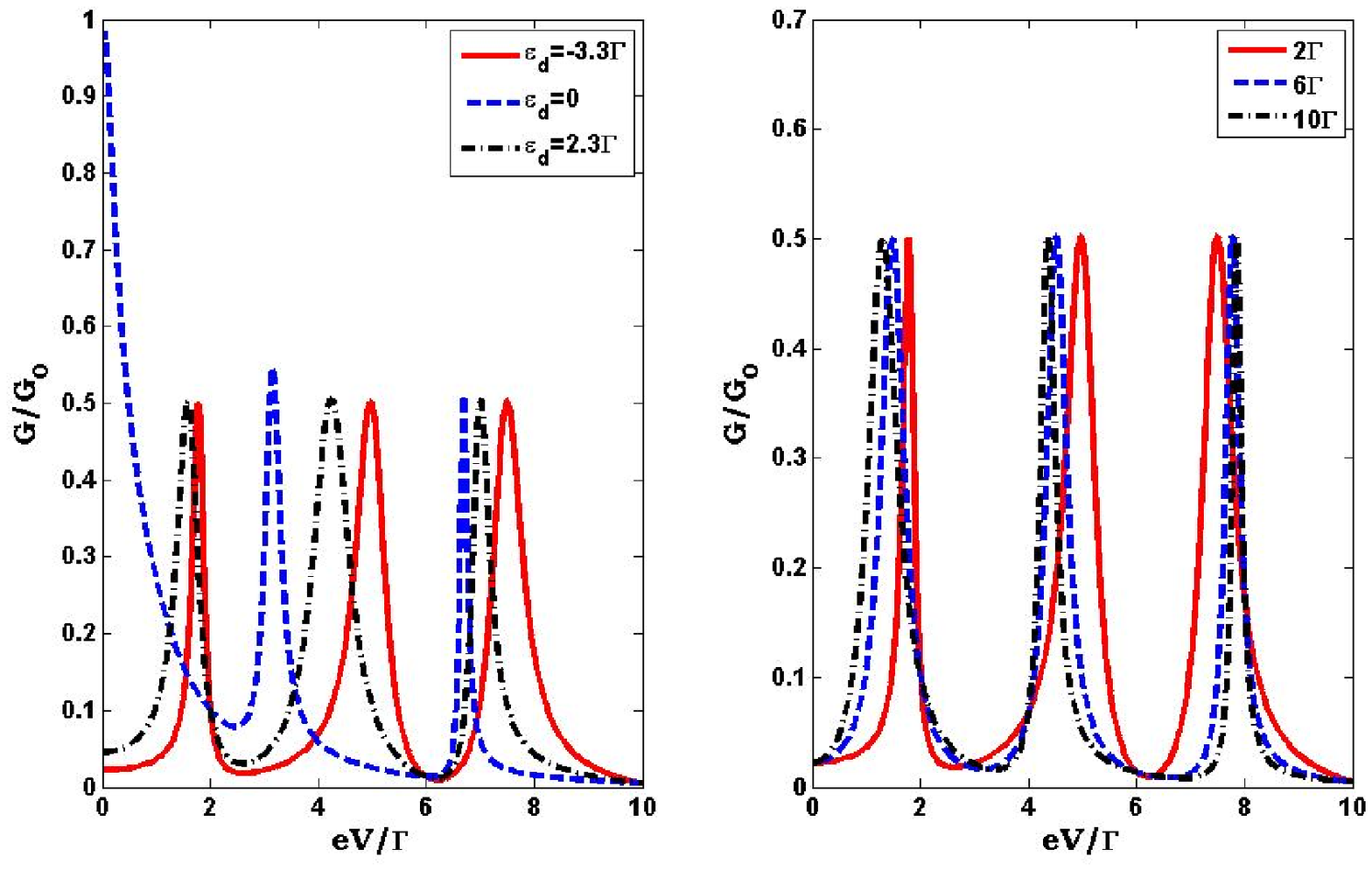}
 \caption{(Color online) The conductance (in units of $G_0$) versus $eV$ (in units of $\Gamma$) with even $N_v$.
 We take $\Gamma_L=\Gamma_R=\Gamma/2=0.5\pi v_M/L$. {\bf Left}: $G$ as a function of $eV$ with $\bar{t}^2/v_M=2\Gamma$
 for different values of $\epsilon_d$. {\bf Right}: $G$ as a function of $eV$ with $\epsilon_d=-3.3\Gamma$ for
 different values of $\bar{t}^2/v_M$.}
 \label{cond3}
\end{center}
\end{figure}

\subsection{The Majorana-fermion representation}

The above results can be understood by introducing the Majorana-fermion representation for the dot electrons:
\begin{eqnarray*}
 \gamma_1=\frac{\xi d+\xi^*d^{\dagger}}{\sqrt{2}} \ , ~~\gamma_2=\frac{\xi d-\xi^*d^{\dagger}}{\sqrt{2}i} \ ,
\end{eqnarray*}
where $\gamma_i$ satisfy the anticommutation relation
\begin{eqnarray*}
 \{\gamma_i,\gamma_j\}=\delta_{ij} \ ,
\end{eqnarray*}
for $i,j=1,2$. We notice that only $\gamma_1$ couples to the chiral Majorana edge states directly. $\gamma_2$
couples the chiral Majorana edge states indirectly through the $\epsilon_d$ term, which represents the hopping
between $\gamma_1$ and $\gamma_2$. The retarded Green functions for $\gamma_i$, which are defined as
\begin{eqnarray*}
 iS_j^r(t_1,t_2)\equiv\Theta(t_1-t_2)\langle\{\gamma_j(t_1),\gamma_j(t_2)\}\rangle \ ,
\end{eqnarray*}
are related to the two-point Green functions of the dot electrons through the following relations:
\begin{eqnarray*}
 S_1^r(t_1,t_2) &=& \frac{1}{2} \! \left[D_r(t_1,t_2)-D_a(t_2,t_1)\right] \\
 & & +\frac{1}{2} \! \left[\xi^2F_r(t_1,t_2)+(\xi^*)^2\tilde{F}_r(t_1,t_2)\right] , \\
 S_2^r(t_1,t_2) &=& \frac{1}{2} \! \left[D_r(t_1,t_2)-D_a(t_2,t_1)\right] \\
 & & -\frac{1}{2} \! \left[\xi^2F_r(t_1,t_2)+(\xi^*)^2\tilde{F}_r(t_1,t_2)\right] ,
\end{eqnarray*}
where
\begin{eqnarray*}
 iF_r(t,t^{\prime}) &\equiv& \Theta(t-t^{\prime})\langle\{d(t),d(t^{\prime})\}\rangle \ , \\
 i\tilde{F}_r(t,t^{\prime}) &\equiv& \Theta(t-t^{\prime})\langle\{d^{\dagger}(t),d^{\dagger}(t^{\prime})\}\rangle
 \ ,
\end{eqnarray*}
are anomalous Green functions for dot electrons. By taking the Fourier transform, we get
\begin{eqnarray}
 S_1^r(\omega) \! \! \! &=& \! \! \frac{1}{2} \! \left[D_r(\omega)-D_a(-\omega)+\xi^2F_r(\omega)+(\xi^*)^2
 \tilde{F}_r(\omega)\right] , \nonumber \\
 S_2^r(\omega) \! \! \! &=& \! \! \frac{1}{2} \! \left[D_r(\omega)-D_a(-\omega)-\xi^2F_r(\omega)-(\xi^*)^2
 \tilde{F}_r(\omega)\right] . ~~~~\label{mdgf12}
\end{eqnarray}
From Eq. (\ref{mdgf12}), we find that
\begin{equation}
 \frac{1}{2}[A(E)+A(-E)]=\pi [\rho_1(E)+\rho_2(E)] \ , \label{mdgf13}
\end{equation}
where $\rho_i(E)=-\frac{1}{\pi}\mbox{Im}[S_i^r(E)]$ with $i=1,2$ are the LDOS's for $\gamma_1$ and $\gamma_2$. That
is, $G(V)$, in fact, measures the sum of the LDOS's for $\gamma_i$.

\begin{figure}
\begin{center}
 \includegraphics[width=0.9\columnwidth]{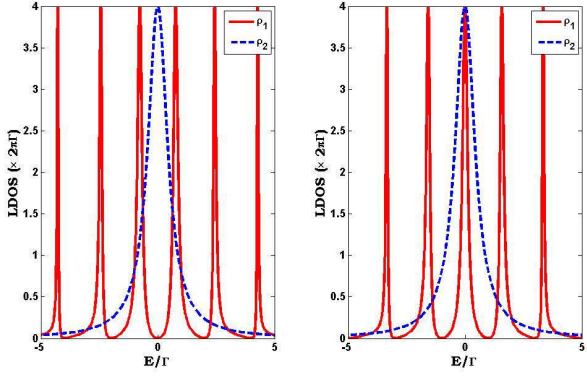}
 \caption{(Color online) LDOS's for $\gamma_1$ and $\gamma_2$ with $\epsilon_d=0$. We take
 $\Gamma_L=\Gamma_R=\Gamma/2=0.5\pi v_M/L$ and $\bar{t}^2/v_M=2\Gamma$. {\bf Left}: odd $N_v$. {\bf Right}: even
 $N_v$.}
 \label{mdos2}
\end{center}
\end{figure}

\begin{figure}
\begin{center}
 \includegraphics[width=0.9\columnwidth]{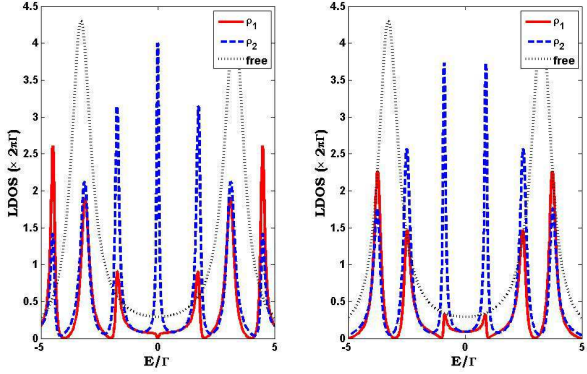}
 \caption{(Color online) LDOS's for $\gamma_1$ and $\gamma_2$ with $\epsilon_d=-3.3\Gamma$. We take
 $\Gamma_L=\Gamma_R=\Gamma/2=0.5\pi v_M/L$ and $\bar{t}^2/v_M=2\Gamma$. {\bf Left}: odd $N_v$. {\bf Right}: even
 $N_v$. For comparison, the dotted line corresponds to the case with $\bar{t}=0$.}
 \label{mdos1}
\end{center}
\end{figure}

The anomalous Green functions for dot fermions can also be obtained by the method of EOM, yielding
\begin{eqnarray}
 F_r(\omega) &=& \frac{(\xi^*)^2M(\omega)}
 {\left[\omega-M(\omega)+\frac{i}{2}\Gamma\right]^2 \! -\epsilon_d^2-[M(\omega)]^2} \ , \nonumber \\
 \tilde{F}_r(\omega) &=& \frac{\xi^2M(\omega)}
 {\left[\omega-M(\omega)+\frac{i}{2}\Gamma\right]^2 \! -\epsilon_d^2-[M(\omega)]^2} \ . \label{mdgf14}
\end{eqnarray}
Inserting Eqs. (\ref{mdgf1}) and (\ref{mdgf14}) into Eq. (\ref{mdgf12}) gives
\begin{eqnarray}
 S_1^r(\omega) &=& \frac{\omega+\frac{i}{2}\Gamma}
 {\left[\omega-M(\omega)+\frac{i}{2}\Gamma\right]^2 \! -\epsilon_d^2-[M(\omega)]^2} \ , \nonumber \\
 S_2^r(\omega) &=& \frac{\omega-2M(\omega)+\frac{i}{2}\Gamma}
 {\left[\omega-M(\omega)+\frac{i}{2}\Gamma\right]^2 \! -\epsilon_d^2-[M(\omega)]^2} \ . \label{mdgf15}
\end{eqnarray}
Hence, the LDOS's for $\gamma_1$ and $\gamma_2$ take the forms:
\begin{eqnarray}
 \rho_1(E) \! \! \! &=& \! \! \! \frac{\Gamma \! \left(E^2+\epsilon_d^2+\Gamma^2/4\right)/2\pi}
 {\left[E^2\! -\! 2EM(E)\! -\! \epsilon_d^2\! +\! \frac{\Gamma^2}{4}\right]^2 \! \! +\! \Gamma^2 \! \! \left\{\epsilon_d^2\! +\! [M(E)]^2\right\}}
 , \nonumber \\
 \rho_2(E) \! \! \! &=& \! \! \! \frac{\Gamma \! \left\{[E-2M(E)]^2+\epsilon_d^2+\Gamma^2/4\right\}/2\pi}
 {\left[E^2\! -\! 2EM(E)\! -\! \epsilon_d^2\! +\! \frac{\Gamma^2}{4}\right]^2 \! \! +\! \Gamma^2 \! \! \left\{\epsilon_d^2\! +\! [M(E)]^2\right\}}
 . \nonumber \\
 & & \label{mdgf16}
\end{eqnarray}
We see that $\rho_i(E)$ with $i=1,2$ are even functions of $E$, i.e. $\rho_i(-E)=\rho_i(E)$, which arises from
the fact that $\gamma_i$ are real fermions. The behaviors of $\rho_i(E)$ with $i=1,2$ are shown in Figs.
\ref{mdos2} and \ref{mdos1}.

For the on-resonance dot, i.e. $\epsilon_d=0$, we notice that $\gamma_2$ does not couple to the chiral Majorana
edge states at all. Thus, its LDOS exhibits a Lorentzian form with a peak at zero energy and width determined by
$\Gamma$, and is independent of the number of vortices in the TSC. On the other hand, due to the coupling to the
chiral Majorana edge states, $\rho_1$ develops several peaks located at the values of $eV/\Gamma$ which are the
real roots of the equation $x=2M(x)$, and the peak values are universal in the sense that
$2\pi\Gamma(\rho_1)_{\mbox{{\tiny max}}}=4$. Hence, the peak values of $G(V)/G_0$ (in units of
$\Gamma_L\Gamma_R/\Gamma^2$) are nonuniversal, except that $G(0)/G_0=2\Gamma_L\Gamma_R/\Gamma^2,
4\Gamma_L\Gamma_R/\Gamma^2$ for odd and even $N_v$, respectively. Moreover, $G(V)/G_0$ approaches the value
$2\Gamma_L\Gamma_R/\Gamma^2$ as $eV/\Gamma\gg 1$ since $\rho_2(E)\rightarrow 0$ when $E/\Gamma\gg 1/2$. On the
other hand, $\rho_1$ becomes zero when $1/M(E)=0$ or $E=E_l$, where $E_l=\frac{2\pi v_M}{L}[l+(N_v+1)/2]$ with
integer $l$ are the quantized energy levels of the chiral Majorana edge modes. This suggests that the coupling to
the chiral Majorana edge states will suppress half of the degrees of freedom for the dot electrons for an
on-resonance dot. One may view this as ``Majorana-fermionization" of half of the degrees of freedom of dot
electrons. In summary, on account of the oscillatory behavior of $\rho_1(E)$, which follows from the coupling to
the chiral Majorana edge states, $G(V)$ for an on-resonance dot becomes an oscillating function of $eV$.

For an off-resonance dot, we notice that the suppression in $\rho_1$ also occurs whenever $E$ matches the
quantized energy levels of the chiral Majorana edge modes. That is, $\rho_1(E)=0$ for $E=E_l$. This implies the
robustness of ``Majorana-fermionization" of dot electrons. Moreover, $\rho_1$ as well as $\rho_2$ are oscillatory
functions of $E$ due to the coupling to the chiral Majorana edge states and $\epsilon_d\neq 0$. In fact, this
oscillation is intimately related to the suppression in $\rho_1$ at $E=E_l$ since it must obey the sum rule:
\begin{eqnarray*}
 \int^{+\infty}_0 \! dE\rho_1(E)=\frac{1}{2} \ .
\end{eqnarray*}
This results in the oscillatory behavior of $G(V)$ as varying $eV$. The peaks in $G(V)$ are located at the values
of $eV/(2\Gamma)$ when they are the real roots of the equation $x^2-2M(x)x=(\epsilon_d/\Gamma)^2\pm 1/4$. Thus,
$G(V)/G_0$ takes the universal value $2\Gamma_L\Gamma_R/\Gamma^2$ at these peaks.

\section{Conclusion}

To sum up, we study the electrical transport through a QD side-coupled to a TSC in the spinless regime. We pay
attention to the behavior of the conductance $G$ as varying the bias $V$. We found that $G$ is an oscillatory
function of $eV$ similar to the one for a multi-level dot. However, the function $G(V)$ in the present case is
distinguished from that for a multi-level dot in three respects. First of all, the former will shift upon adding
or removing a vortex in the TSC, while such an effect is not observed for the latter. Next, for an off-resonance
dot, the value of the conductance peak in the former case is only half of the one in the unitary limit, whereas
for the latter, it will reach a non-universal value depending on the energy levels in the dot as long as $eV/2$
matches one level in the dot. Finally, for an on-resonance dot, the conductance peak in the former case will
approach a universal value $G/G_0=2\Gamma_L\Gamma_R/\Gamma^2$ at large bias. In general, there is no such a
behavior for the latter. We consider these features the signatures of chiral Majorana edge states.

The oscillatory behavior of $G(V)$ can be understood by introducing the Majorana representation of dot electrons.
The dot electron is composed of two Majorana fermions. Only one of them is coupled to the chiral Majorana edge
states directly. We show that this coupling results in the suppression in the LDOS whenever the energy matches
one of the quantized energy levels of the chiral Majorana edge states. This phenomenon is dubbed as Majorana
fermionization because only the other Majorana fermion, which is not directly coupled to the chiral Majorana edge
states, survives at these energies. It is interesting to explore similar phenomena in other situations.

Finally, we would like to emphasize that the above results are obtained assuming zero temperature and without
other possible dissipations. Extension to the finite temperature can start with Eq. (\ref{i1}). Dissipations
arising from the environment, however, involve the change of the model. Dissipation effects may suppress the
tunneling rate or cause a nontrivial phase diagram such that the results obtained from the tunneling spectroscopy 
to identify the signature of the chiral Majorana liquid may be dubious. The effects of ohmic dissipations on the 
tunneling between the normal metallic lead and the end of a $1$D TSC has been studied,\cite{DL} which shows 
distinct temperature behaviors for the zero-bias conductance peaks due to the Majorana fermion end mode and other 
effects. It deserves to include the dissipation effects into our analysis to see how the results we obtained are 
modified.

\acknowledgments

The author would like to thank Y.-W. Lee and C.S. Wu for enlightening discussions.


\end{document}